\documentclass[10pt,letterpaper,twocolumn]{article} 
\usepackage{ol}

\usepackage{hyperref}
\usepackage{amsmath}

\begin{document}

\twocolumn[ 

\title{Absorption-induced confinement of lasing modes in diffusive random medium}

\author{A. Yamilov, X. Wu, and H. Cao}

\address{Department of Physics and Astronomy, Northwestern University, Evanston, IL 60208}

\author{A. L. Burin}

\address{Department of Chemistry, Tulane University, New Orleans, LA 70118}

\begin{abstract}
We present a numerical study of lasing modes in diffusive random media with local pumping. The reabsorption of emitted light suppresses the feedback from the unpumped part of the sample and effectively reduces the system size. The lasing modes are dramatically different from the quasimodes of the passive system (without gain or absorption). Even if all the quasimodes of a passive diffusive system are extended across the entire sample, the lasing modes are still confined in the pumped volume with an exponential tail outside it. The reduction of effective system volume by absorption broadens the distribution of decay rates of quasimodes and facilitates the occurrence of discrete lasing peaks.  
\end{abstract} 



] 

Over the past few years, there have been many studies on random lasers with coherent feedback \cite{cao_review}. In a strongly scattering active medium, recurrent scattering events could provide resonant feedback for lasing. With sufficient gain, lasing oscillation might occur at discrete frequencies that are determined by the interference of scattered light. In our initial experiments with highly disordered semiconductor powder and polycrystalline films, the transport mean free path was short, leading to small laser cavities \cite{PRL00}. Our experimental observation was well reproduced in the numerical simulation of lasing in localized modes \cite{soukoulis,sebbah,burin}.  However, lasing with coherent feedback was realized also in weakly scattering random media \cite{frolov,ling,polson1}.  Tight focusing of pump light was necessary to observe discrete lasing peaks, namely, the pump beam was focused to a region of size much smaller than the entire sample. Imaging of laser light on the sample surface revealed that the lasing modes were not extended over the entire random medium, instead they were located inside the pumped region with an exponential tail outside it \cite{PRE02}. Since the quasimodes of a random system far from the onset of localization were usually extended states, the lasing modes were regarded as some types of anomalously localized states, either almost localized states \cite{apalkov,polson2}, or prelocalized states \cite{mirlin,PRE02}. 

The anomalously localized states should be rare in the diffusive samples. Yet no matter where on the sample the pump beam is focused, we always observe lasing modes that are spatially confined in the pumped region. Moreover, the lasing threshold does not fluctuate much as the pump spot is moved across the random medium. We believe the contradiction to the theory of anomalously localized states originates from the assumption that lasing occurs in the quasimodes of the passive random medium. In this letter, we will demonstrate with numerical simulation that this assumption is not valid when absorption at the emission wavelength is significant outside the pumped volume. The reabsorption of emitted light suppresses the feedback from the unpumped part of the sample and effectively reduces the system size. The lasing modes are formed by the interplay between gain and absorption, and therefore, are dramatically different from the quasimodes of the passive system (without gain or absorption). Even if all the quasimodes of a passive diffusive system are extended across the entire sample, the lasing modes are still confined in the pumped volume with an exponential tail outside it. 

We use the finite-difference time-domain (FDTD) method to simulate lasing in the transverse magnetic (TM) modes of two-dimensional (2D) random media. The disordered system is a collection of dielectric cylinders placed at random in vacuum. The diameter of the cylinders is $160$nm, the refractive index $n=2$, and the filling fraction is about $50\%$. The total size of the system is 9.2$\mu$m $\times$ 9.2$\mu$m. The random medium is surrounded by vacuum, which is terminated by uniaxial perfectly matched absorbing layer\cite{taflove}.  The wavelength of interest is around 650nm. To verify light transport in the 2D system is diffusive, we calculate the continuous wave (CW) transmission through a slab of the random medium with  the thickness $L$ = 9.2$\mu$m. The ensemble-averaged intensity profile $\langle I(z)\rangle$ is fitted well with the diffusion expression $(L+z_0-z)/(L+2z_0) $\cite{vanrossum}, where $z$ is the coordinate normal to the slab. The extrapolation length $z_0$ is related to the transport mean free path $\ell$ as $z_0/\ell=\pi/4 (1+R)/(1-R)$. $R$ is angularly-averaged boundary reflection coefficient, which accounts for interface effects \cite{vanrossum}. Using the expression for $R$ in the 2D system \cite{gilbert_thesis}, we estimate $z_0/\ell\simeq 1.85$. The value of $z_0$ is extracted from the fitting of $\langle I(z)\rangle$, that gives $\ell\simeq 1.3\mu m$. The small value of $\ell/L \sim 0.12$ as well as the large value of $k\ell \sim 13$ ensures the 2D random system of lateral dimension $L= 9.2 \mu$m is in the diffusive regime. 

\begin{figure}
\centerline{\rotatebox{-90}{\scalebox{0.35}{\includegraphics{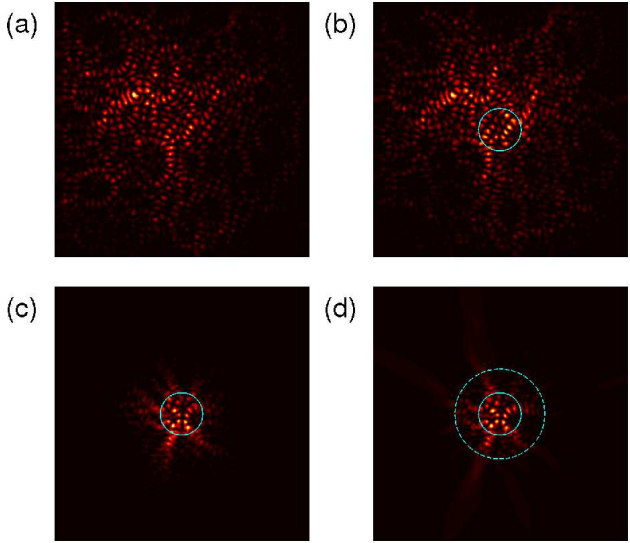}}}}
\vskip-0.1cm
\caption{\label{profiles} Spatial intensity distribution of (a) the quasimode with the longest lifetime in a passive diffusive system; (b) the (first) lasing mode with gain inside the circular region near the center and no absorption outside it; (c) the (first) lasing mode with gain inside the circular region near the center and absorption outside it; (d) the (first) lasing mode with random medium beyond one $\ell_p$ (dashed circle) removed. }
\end{figure}

\begin{figure}
\centerline{\rotatebox{-90}{\scalebox{0.33}{\includegraphics{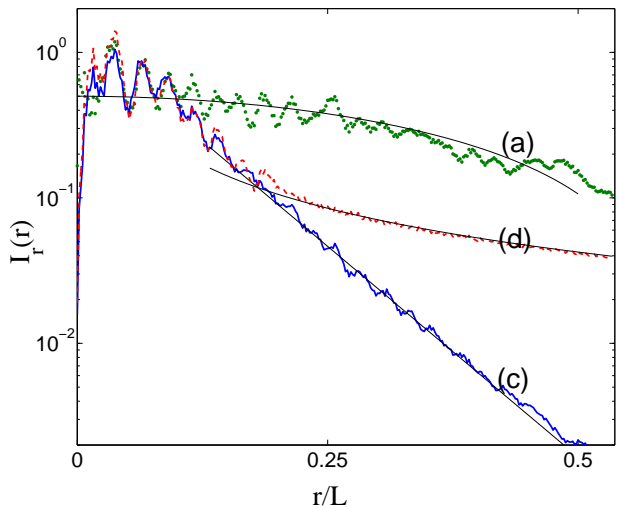}}}}
\vskip-0.1cm
\caption{\label{radial} Radial dependence of the angularly-integrated intensities of the modes (a,c,d) shown in Fig. \ref{profiles}. (a) is compared to the diffusive mode profile; the tail of (c) is fitted with an exponential decay; the falloff of (d) outside the random medium is compared to $r^{-1}$.}
\vskip-0.5cm
\end{figure}

We start with a study of the quasimodes (eigenmodes of the Maxwell's equations) in the passive random medium. A short excitation pulse, whose spectrum is centered at $\lambda_0\simeq 650$nm, is launched in the center of the sample. After the excitation pulse is gone, the total electromagnetic energy stored inside the random system $U(t)$ exhibits a non-exponential decay in time as a result of multi-mode excitation. However, after a sufficiently long time, $U(t)$ changes to an exponential decay because only one mode with the longest lifetime is left inside the system. The exponential decay rate of $U(t)$ is equal to the decay rate $\gamma$ of this quasimode.  For example, for one realization shown in Fig. \ref{profiles}(a), the Fourier transform of electric field $E(t)$ gives the wavelength of the quasimode $\lambda = 646$nm. The spatial profile of the electric field becomes stable in time, and it reflects the wavefunction of this longest-lived quasimode. As shown in Fig. \ref{profiles}(a), this quasimode is extended across the sample. When we remove some cylinders at the sample boundary, e.g., the cylinders outside a circle of radius $L/2$ from the mode center, the mode profile changes dramatically. The sensitivity of the quasimode to the boundary confirms that it is an extended state.  Fig. \ref{radial} shows the radial dependence of the angularly-integrated intensity, $I_r(r)= \int |E(r, \theta)|^2 r d \theta$, where $r$ is the radial coordinate, $\theta$ is the polar angle, $E(r, \theta)$ is the electric field distribution of the quasimode. For comparison, the radial profile of the lowest diffusion mode $\cos{\left[\pi x/(L+2z_0)\right]}\cos{\left[\pi y/(L+2z_0)\right]}$ is also plotted in Fig. \ref{radial}. It describes $I_r(r)$ relatively well.


Now we introduce gain and reabsorption into the random medium. In the previous studies \cite{sebbah,soukoulis}, the gain medium is modeled as four-level atomic system where the lasing transition is from the third level to the second level. In the absence of pumping, all the atoms are assumed to be in the first level. Since the electronic population in the second level is zero, there is no absorption at the lasing wavelength in the unpumped region. To simulate spatially non-uniform gain and reabsorption of the laser emission, we use the semi-classical Lorentz model \cite{taflove}. The linear gain/absorption is modeled by negative/positive conductance \cite{PRL00}. By introducing negative conductance to the pumped region and positive conductance to the unpumped region, we are able to describe both light amplification inside the pumped region and reabsorption of the emitted light outside the pumped region. More specifically, the cylinders have the conductance\cite{taflove}
\begin{equation}
\sigma(\omega)=\frac{\sigma_0/2}{1+i(\omega-\omega_0)T_2}+\frac{\sigma_0/2}{1+i(\omega+\omega_0)T_2}.
\label{sigma}
\end{equation}
The sign of $\sigma_0$ determines light is amplified or absorbed, whereas the amplitude of $\sigma_0$ sets the magnitude of gain/absorption. $\omega_0$ and $1/T_2$ determine the center frequency and width of the gain/absorption spectrum, respectively. The absence of gain saturation in Eq. \ref{sigma} is not crucial in our simulation, as our goal is to find the first lasing mode at or slightly above the lasing threshold. A seed pulse is launched at $t=0$ to initiate the amplification process.   The lasing threshold is defined by the minimum gain coefficient ($-\sigma_0$) at which the electromagnetic energy stored inside the random system grows in time and eventually diverges. 

We first consider the case of local pumping with no absorption outside the pumped region. We use the same random sample as in Fig. \ref{profiles}(a), and introduce gain to the central part marked by the circle in Fig. 1(b). The gain spectrum is centered at 650nm and has a width of 52nm. The lasing mode shown in Fig. \ref{profiles}(b) is identical to the quasimode of the passive system in Fig. \ref{profiles}(a).  Although optical gain is concentrated within the circle near the center, the lasing mode is extended throughout the entire sample. As we reduce the pump area by decreasing the radius of the circle, the lasing threshold is increased, but the lasing mode profile remains the same. This result indicates that the lasing mode in a diffusive random medium is the extended quasimode of the passive system, even when the pumped region is smaller than the mode size.  

However, the above statement is valid only when there is no absorption outside the pumped region, which is not the case in most experiments. For example, Rhodamine dye, which is widely used to provide gain for random lasers, has significant overlap between its absorption band and emission band. Therefore, photons that are emitted by the excited rhodamine molecules inside the pumped region may diffuse into the surrounding unpumped region and be absorbed by the rhodamine molecules there. The absorption reduces the probability of light returning to the pumped region, thus suppresses the feedback from the unpumped region. To simulate the reabsorption, we introduce absorption outside the pump area. The bulk absorption length (without scattering) $\ell_a = 1.06 \mu$m. Fig. \ref{profiles}(c) shows the lasing mode profile, which is very different from that in Fig. \ref{profiles}(a) or \ref{profiles}(b).  The wavelength of the lasing mode also differs by about $4$nm.  Therefore, the lasing mode in the presence of reabsorption is a new mode, completely different from the quasimode of the passive system. Due to reabsorption outside the pump area, the lasing mode is confined more or less inside the pumped region. The radial profile of the lasing mode, shown in Fig. \ref{radial}, features an exponential decay outside the pumped region. The decay length, $\sim 0.8\mu$m, is equal to the diffusive absorption length $\ell_p = (\ell_a \ell /2)^{1/2}$. To confirm this result, we simulate transmission of CW plane wave through a slab of disordered absorbing medium, and obtain the same attenuation length. 

Performing calculations similar to those shown in Figs. \ref{profiles},\ref{radial} for many disorder configurations, we find that the observed effect is indeed general. The reabsorption, which suppresses the feedback from the unpumped part of the sample, effectively reduces the system size. To check this conjecture, we remove all the random medium beyond one $\ell_p$ from the pump area [dashed circle in Fig. \ref{profiles}(d)], and repeat the calculation. The frequency and spatial profile of the lasing mode remain the same  [Fig. \ref{profiles}(d)], despite the drastic change of the random system. The radial profile of the lasing mode outside the random medium (Fig. \ref{radial}) exhibits a trivial $r^{-1}$ dependence. This result indicates the lasing mode is an  extended state within the effective volume, $V_{eff}$, that is made of the optically active region plus the buffer layer of thickness $\ell_p$ around it. 

The reduction of the effective system volume leads to a decrease of the Thouless number $\delta \equiv \delta\nu / \Delta\nu$, where $\delta\nu$ and $\Delta\nu$ are the average mode linewidth and spacing respectively. In a 3D diffusive system $\delta \nu \propto V_{eff}^{-2/3}$ and $\Delta \nu \propto V_{eff}^{-1}$, therefore, $\delta \propto V_{eff}^{1/3}$.  The smaller the value of $\delta$, the larger the fluctuation of the decay rates $\gamma$ of the quasimodes  \cite{mirlin,chabanov}. The variance of the decay rates\cite{mirlin} $\sigma^2_{\gamma}=\langle \gamma\rangle^2/\delta$, where the average decay rate $\langle \gamma\rangle\sim D/V_{eff}^{2/3}$. We believe the broadening of the decay rate distribution along with the decrease of the total number of quasimodes (within $V_{eff}$) is responsible for the observation of discrete lasing peaks in the tight focusing condition. Despite its value is reduced, the effective Thouless number is still much larger than 1 due to weak scattering. As a result, the lasing modes are the extended states within the effective volume. Because $\sigma_{\gamma}/\langle \gamma\rangle\ll 1$, the minimum decay rate is still close to $\langle \gamma\rangle$, leading to relatively small fluctuation of lasing threshold \cite{burin2,patra,yamilov}.

The authors acknowledge Prof. A. Z. Genack, Drs. C. Vanneste and P. Sebbah for stimulating discussions. This work is supported by the National Science Foundation under Grant No. DMR-0093949. AB acknowledges the support from the Louisiana Board of Regents RCS, Project 117A. Authors electronic address: a-yamilov@northwestern.edu; h-cao@northwestern.edu.

\end{document}